**Searching for Shakespeare in the Stars**

Eric Lewin Altschuler

School of Medicine

University of California, San Diego

9500 Gilman Drive, 0606

La Jolla, CA 92093-0606

**Abstract**

The question of the authorship of Shakespeare's plays has long been debated. The two leading contenders are W. Shakspere (1564-1616) and Edward de Vere the 13[th] Earl of Oxford (1550-1604). Here we note that Shakespeare's references to important events and discoveries in astronomy and geophysics in 1572 and 1600, but not to similarly important events of 1604, 1609 and 1610, especially given Shakespeare's frequent references to and knowledge of the physical sciences, might be able to shed some light on the authorship question.

"After four centuries, Shakespeare remains the most haunting of authors," Joseph Sobran has written in a new book[1]. "He seems to know us better than we know him." As time has gone on, doubt that the actor W. Shakspere (1564-1616) wrote the Shakespeare plays has increased. Furthermore, it should be noted that biographies of Shakespeare did not begin to take shape until nearly a century after Shakespeare's death. Numerous individuals have been proposed as the author of the Shakesperean plays including Francis Bacon (1561-1626), Christopher Marlowe (1564-1593), and even Queen Elizabeth I. However, the most serious contender still left standing is Edward de Vere the 13th Earl of Oxford (1550-1604), himself an actor and poet, and the founder of Shakspere's acting company the Lord Chamberlain's Men. Summarizing and strengthening previous arguments and adding many of his own, Sobran makes perhaps the strongest case ever for Oxford as Shakespeare. For example: Shakespeare used many sources for his plays, but none may have dated later than 1603; plays written by others with the name Shakespeare started appearing in 1605 after Oxford's death, but while Shakspere was still alive; many, including playwrite Ben Johnson, during the years 1604-1616, seem to speak of Shakespeare as being dead; there is a confluence of events in the Shakespearean plays with the life of Oxford but not that of Shakspere; the young man to whom many of the Sonnets (~1593-1600) were addressed appears to have been Henry Wriothesley, Earl of Southampton, a more unlikely addressee for Shakspere than for Oxford; the Sonnet's, which even Stratfordians (partisans for Shakspere as Shakespeare) concede were written in the voice of an old man, were written at a time when Shakspere was in his early thirties, while Oxford was in his fifties and ailing. Of course, these literary arguments in favor of Oxford as Shakespeare are disputed mightily by the Stratfordians. While some may believe it doesn't matter who Shakespeare was, the question has intrinsic interest, and, as well, the vast amount of informal and professional study of Shakespeare might be enhanced by knowledge of his identity. Here I note that Shakespeare's references to important events and discoveries in astronomy and geophysics in 1572 and 1600, but not to similarly important events of 1604, 1609 and 1610, especially given Shakespeare's frequent references to and knowledge of the physical sciences[2], might be able to shed some light on the authorship question.

Shakespeare was remarkably learned in numerous fields including medicine[3], law[4], and the physical sciences[2]. With regards to astronomy, for example, Shakespeare's characters in *Henry IV Part I* (act 2, scene 1) use the position of the constellation Ursa Major to tell the hour of the night. Now, at the beginning of *Hamlet* Bernardo says, "Last night of all,/When yond same star that's westward from the pole/Had made his course to illume that part of heaven/Where now it burns, Marcellus and myself,/The bell then beating one." Other events in the play suggest that the star was sighted in November (see ref. 5 and refs. therein). What might be the identity of the bright star in the skies of Denmark west of the pole at 1AM on a November evening during the perturbed times in that Kingdom when *Hamlet* took place? Recently, Olson, Olson and Doescher have studied this question[5]. They could find no particularly bright fixed star visible in the sky in Denmark in November, west of the pole, at 1AM. However, they have made the fascinating observation that the "new" star which appeared in the sky in November 1572 in the constellation Cassiopeia, now known to be a supernova (SN1572A), does fulfill all these criteria[5]! The new star was described in detail by Danish astronomer Tycho Brahe (1546-1601)[6]. The notion that this star in *Hamlet* might be SN1572A is consistent with an earlier observation that the names 'Rosenkrans' and 'Guldensteren' are among those appearing in a portrait of Brahe surrounded by coats-of-arms of his ancestors[7]. In 1572 Oxford was twenty-two while Shakspere was only eight. Perhaps the memory of the new star was etched into the memory of eight year-old Shakspere, but more likely into that of twenty-two year-old Oxford especially as in England it was Lord Burghley, Oxford's father-in-law, whom Queen Elizabeth asked to investigate the new heavenly development.

In 1577 a great comet was visible in Europe, and there would not be another until 1607 (an appearance of Haley's comet). Interestingly Shakspeare often refers to comets in plays such as *Hamlet* (1600-1601) and *Henry VI Part 1* (1589-1592) plays which appeared before 1607. Perhaps the comet of 1577 was imprinted in the mind of the thirteen year old Shakspere, but again it is more likely to have been so for Oxford.

In 1600 in a landmark book British physician and naturalist William Gilbert proposed the idea that the earth may have a magnetic field[8]. Shakespeare seemed aware of this theory: in *Troilus and Cressida* (1601-1602) he writes, (III.2.184-186) "As true as steel, as plantage to the moon,/As sun to day, as turtle to her mate,/As iron to adamant, as earth to the centre." In the same play, (IV.2.109-111) "But the strong base and building of my love/Is as the very centre of the earth,/Drawing all things to it." Thus, Shakespeare knew, apparently about recent developments in science including SN1572A and also Gilbert's theory of geomagnetism. What about Shakespeare's knowledge of later astronomical events and discoveries?

In October 1604 near a conjunction of Mars, Saturn and Jupiter another new star was seen in the sky (SN1604A). Shakespeare makes no mention of SN1604A. We might have expected Shakespeare to notice SN1604A given the observation "Saturn and Venus in conjunction!" in *Henry IV Part 2* (II.4.286).

Consider the following from *Henry VI Part 1* (I.2.1-2) "Mars his true moving, even as in the heavens/So in the earth, to this day is not known." This would appear to be a reference to the fact that the orbit of Mars was not well understood, even seen to be going backward according to some models of its orbit proposed by the time of *Henry VI Part 1* (1589-1592). In 1609 in his *Astronomia Nova*[9] Kepler discussed his first two laws of motion, and, in particular gave the first proper account of the orbit of Mars. So, Shakespeare seems aware of the problems previous to 1609 of explaining the orbit of Mars, but despite the current dating of five plays after 1609 *Cymbeline* (1609-1610), *Winter's Tale* (1610-1611), *The Tempest* (1611), *Henry VIII* (1612-1613), and *Two Noble Kinsmen* (1613), Shakespeare never felt it important to mention the resolution of the confusion over the orbit of Mars. (The standard dating for Shakespeare's plays is not based on publication, many of which were not published until 1623, but on performance dates and other evidence.)

In 1609 based on early reports of the telescope, Galileo Galilei fashioned his own and turned it toward the heavens. Among the discoveries reported in his 1610 book[10] were sunspots, the phases of Venus (similar to the phases of the Earth's moon), hills, valleys and other imperfections on the surface of the Moon, and the moons of Jupiter. None of these discoveries are mentioned in Shakespeare. Shakespeare's omission of these discoveries seems striking given his substantial knowledge and frequent mention of such topics. For example: Shakespeare was a most keen observer of the sun, discussing *in Henry VI Part 3* (II.1. 25-32) the phenomenon of parhelia ("mock suns") which seem to appear near the real Sun due to ice crystals. "*Edward*: Dazzle mine eyes, or do I see three Suns?/ *Richard*: Three glorius suns, each one a perfect sun;/Not separated with the racking clouds,/But sever'd in a pale clear-shining sky./See, see! They join, embrace, and seem to kiss,/As if they vow'd some league inviolable:/Now are they but one lamp, one light, one sun./In this the heaven figures some event." As well, Shakespeare refers to the Sun more than forty times in the late plays, but never to sunspots. Shakespeare appreciates the brightness of Venus (*Midsummer Night's Dream* III.2.61), and that Venus can be an evening (*All's Well that Ends Well* II.1.166-167) or morning star as mentioned/See, see! They jo*in, embrac*e, and seem to kiss,/As if they vow'd some league inviolable:/Now are they but one lamp, one light, one sun./In this the heaven figures some event." As well, Shakespeare refers to the Sun more than forty times in the late plays, but never to sunspots. Shakespeare appreciates the brightness of Venus (*Midsummer Night's Dream* III.2.61), and that Venus can be an evening (*All's Well that Ends Well* II.1.166-167) or morning star as mentioned in the late play *Henry VIII* (III.2.366-367, 371-372). Shakespeare has more than fifteen mentions to the Moon in the late plays without mention of the imperfections on its surface, and more than ten references to Jupiter without mention of its moons.

In conclusion, Shakespeare's works show us that the instrument he was using to examine the Heavens was the human eye—indeed a most keen and learned eye—but not a telescope: Shakespeare knew about SN1572A, and Gilbert's discussion of geomagnetism in 1600, but apparently not about SN1604A, sunspots, the phases of Venus, the imperfections on the surface of the Moon, or the moons of Jupiter. There are many possible explanations why Shakespeare did not write about any of these topics, however, the most parsimonious is that the Bard was not alive to know of these new developments in astronomy. This would be consistent with Oxford (1550-1604), but not Shakspere (1564-1616) as Shakespeare.